\documentclass{ifacconf}
\usepackage{kpfonts}
\usepackage{times}
\usepackage{amsmath} % assumes amsmath package installed
\usepackage{amssymb}  % assumes amsmath package installed
\usepackage{mathrsfs} 
\usepackage{graphicx}
\usepackage{enumerate}
\usepackage{comment}
\usepackage{algorithm}
\usepackage{algpseudocode}
\usepackage{tcolorbox}
\usepackage{tabularx}
\usepackage{array}
\usepackage{dsfont}
\newtheorem{ass}{Assumption}

\usepackage{booktabs}

\usepackage{setspace}
\usepackage{caption}
\newcommand*{\QEDA}{\null\nobreak\hfill\ensuremath{\blacksquare}}

\usepackage{natbib}   % required for bibliography
\begin{document}
\begin{frontmatter}

\title  {Data-Driven Robust Safety Verification for Markov Decision Processes\thanksref{footnoteinfo}} 

\thanks[footnoteinfo]{This work has been supported by the Independent Research Fund Denmark (DFF) under Project SAFEmig (Project Number 3105-00173B).}

\author[]{Abhijit Mazumdar,} 
\author[]{Manuela L. Bujorianu, and}
\author[]{Rafal Wisniewski} 

%\address[] { Department of Computer Science, University College London, England, UK (e-mail: l.bujorianu@ucl.ac.uk),}
\address[]{Section of Automation \& Control, Aalborg University, 9220 Aalborg East, Denmark (e-mail: abma, lmbu, raf @es.aau.dk)}

\begin{abstract}                
In this paper, we propose a data-driven robust safety verification framework for stochastic dynamical systems modeled as Markov decision processes with time-varying and uncertain transition probabilities. Rather than assuming access to the exact nominal transition kernel, we consider the realistic setting where only samples from multiple system executions are available. These samples may correspond to different transition models inside an \textit{ambiguity set} around the nominal transition kernel. Using these observations, we construct a unified ambiguity set that captures both inherent run-to-run variability in the transition dynamics and finite-sample statistical uncertainty. This ambiguity set is formalized through a Wasserstein-distance ball around a nominal empirical distribution and naturally induces an interval Markov decision process representation of the underlying system. Within this representation, we introduce a \textit{robust safety function} that characterizes reach-avoid type probabilistic safety under all transition kernels consistent with the interval Markov decision process. We further derive high-confidence safety guarantees for the true, unknown time-varying system. A numerical example illustrates the applicability and effectiveness of the proposed approach.
\end{abstract}


\end{frontmatter}
%===============================================================================
\section{Introduction}
Safety verification is vital to safety-critical systems, where unmodeled uncertainties and unpredictable variations can pose serious risks. Traditional approaches to probabilistic safety verification often assume that the system’s behavior can be precisely characterized by known probability distributions. However, accurately identifying these probability distributions in practical scenarios is often challenging due to limited data, environmental variability, or complex dynamics. Consequently, any safety guarantees based on approximate or estimated probability models may become invalid if there is a mismatch between the assumed and actual distributions governing the system's transitions.
\par In stochastic optimization, distributionally robust optimization (DRO) has emerged as a powerful tool for providing reliable guarantees under model uncertainty. DRO frameworks account for this uncertainty by optimizing for the worst-case distribution within a specified ambiguity set which is a collection of probability distributions that captures possible variations around the nominal model. By considering the worst-case scenario within this set, DRO ensures that the safety guarantees remain robust, even when the true distribution deviates from the computed or assumed one. This approach allows safety-critical systems to operate with greater resilience to uncertainties in their underlying probabilistic models. There are many approaches used in DRO such as Wasserstein distance-based \citep{mohajerin2018data,gao2023distributionally,yu2024fast}, moment-based \citep{hao2021moment}, $\Phi$-divergence-based \citep{wu2024understanding,love2014classification}, etc.

\textit{Related Literature:}
The concept of safety that we consider in this work is the so-called $p$-safe. It has been developed in \cite{wisniewski2020p} and extensively studied in \cite{wisniewski2021safety, bujorianu2021stochastic, bujorianu2021p, bujorianu2020stochastic}. A system is called to be $p$-safe if the system states do not visit the dangerous states before reaching the goal states with a probability more than a given threshold $p$. In \cite{wisniewski2023probabilistic}, considering a Markov decision process (MDP), the safety function is reformulated by expected cost and a recursive expression is presented using dynamic programming problem.
%Assume that the probability distribution of uncertain variables are partially unknown, distributionally robust optimization solves a minimax problem, which finds the optimal solution under the worst case.
\par In solving DRO optimization, the ambiguity set plays an important role. Wasserstein metric, which is a distance on the space of probability distributions, has been widely used to construct ambiguity set. However, use of Wasserstein distance makes the original problem an infinite-dimensional optimization over probability measures. By using Kantorovich duality \citep{mohajerin2018data,kantorovich1958space} resolves the infinite-dimension issue in the worst-case problem and represents the worse-case problem by a finite-dimensional convex optimization. In the context of an MDP, \cite{iyengar2005robust} introduces a distributionally robust Bellman equation based on which, a robust value iteration algorithm and a robust policy iteration algorithm are developed. Considering Wasserstein ambiguity set, \cite{yang2017convex} presents a method for finding the robust optimal policy. To carry out safety assessment, \cite{yang2018dynamic} uses moment-based ambiguity to define the uncertain distribution, enabling safety verification under partially known disturbance distributions. However, these works only consider either model ambiguity or ambiguity due to finite data samples, not both at the same time.
\\
\\
\textit{Our Contributions:}
In this paper, we study probabilistic safety for stochastic dynamical systems whose behavior may vary over time. We model the system as an \textit{interval Markov decision process (IMDP)} with transition dynamics that are not fixed but can change slightly from one system execution to another due to disturbances, unmodeled effects, or shifts in operating conditions. Since the true transition dynamics are unknown, we rely on data collected from multiple runs, where each run provides one observation of the successor state. These samples are not required to come from the same underlying model, but we assume that the run-to-run variability remains within a bounded range. This leads to two sources of uncertainty in the data: \textit{inherent variability in the system dynamics} and \textit{statistical error from having only a finite number of observations}. We develop an approach that simultaneously accounts for both uncertainties and provides safety guarantees that remain valid under such model variation.

Specifically, our goal is to compute the worst case probability of reaching unsafe states of the IMDP before visiting goal states. To do so, we formulate the problem as a distributionally robust sequential optimization. We use Wasserstein distance to construct the ambiguity set \citep{mohajerin2018data}. To characterize distributionally robust safety, we  introduce the robust safety function which is the worst case probability of reaching unsafe states before visiting goal states. We observe that the robust safety function is the solution of distributionally robust optimization with Wasserstein ambiguity set. 
\par The following are our main contributions:
\begin{enumerate}[i)]
  %  \item We introduce a new safety analysis framework for an interval MDP with both time-dependent and run-dependent transition probabilities.
  %  \item Unlike the existing work on interval MDPs, we model uncertainty using a dual-source structure that captures both inherent variability in the transition dynamics and statistical error due to finite data. 
   % \item By constructing a data-driven confidence region that simultaneously accounts for model variation and empirical uncertainty, we obtain high-confidence safety guarantees for the underlying interval MDP. To the best of our knowledge, our result provides the first high-confidence safety assessment results for interval MDPs.
    %\item While existing data-driven DRO approaches assume that all data samples are generated by a fixed distribution, we relaxed this assumption. We consider the case where the data samples are generated by independent but different probabilities inside an ambiguity set.
\item In contrast to existing IMDP formulations that treat uncertainty as arising from either model variability or finite-sample statistical uncertainty, we introduce a \emph{dual–source uncertainty model} that explicitly separates $(a)$ inherent variability in the true transition dynamics and $(b)$ statistical uncertainty induced by finite data. This decomposition enables a finer and more realistic characterization of uncertainty.

\item Considering the dual–source structure, we construct a \emph{data-driven confidence region} for the transition kernel that jointly accounts for both model variability and sampling error. This yields high-confidence upper bounds on the reach–avoid safety probability. To the best of our knowledge, this constitutes the first high-confidence safety certification method for IMDPs.

\item Existing data-driven distributionally robust optimization (DRO) methods implicitly assume that all observed samples are drawn from a single, fixed underlying transition distribution. We relax this restrictive assumption by allowing observations to be generated by independent but \emph{non-identically distributed} probabilities, each lying within the prescribed ambiguity set. This generalization enables safety analysis in environments with run-to-run variation and dynamic uncertainty.

\end{enumerate}
The organization of the rest of the paper is as follows. In section \ref{problem_formulation}, we introduce the MDP and various definitions. We introduce Wasserstein metric to describe the distance of two probability distributions and use Kantorovich’s duality to reformulate it. Then, we formally define robust $p$-safety. We present the main results of the paper in Section \ref{DR safety verification}.
We provide an upper bound for the robust safety function for the true interval MDP in terms of the empirical robust safety function. In section \ref{numerical results}, we demonstrate our main result using a numerical example. Finally, we present the concluding remark and our future work plan in section \ref{sec_conclusion}.

\textit{Notations:}
For any set $S$, we denote by $\Delta(S)$ the set of all probability measures on $S$. 
For a real number $x$, $|x|$ denotes its absolute value, whereas for a set $S$, $|S|$ denotes its cardinality. 
Given a probability measure $\mu$ over a discrete set $X$, the space $L^{1}(\mu)$ consists of all functions $f$ satisfying $\sum_{x\in X} |f(x)|\,\mu(x) < \infty$. For a function $G : S \to \mathbb{R}$, the expression $\{G(i)\}_{i\in S}$ represents a vector of dimension $|S|$ with components indexed by $S$. 
For real numbers $r_1$ and $r_2$, the notation $[r_1, r_2]$ denotes the standard closed interval. 
For integers $k_1 < k_2$, the notation $[k_1 : k_2]$ denotes the set of integers in the interval $[k_1, k_2]$. 
For a vector $h \in \mathbb{R}^n$, the $\ell_1$-norm is written as $\|h\|_1$. 
For an event $T$ in an experiment, $\mathbb{P}[T]$ denotes the probability that the event $T$ occurs. 
The symbols $\prod$ and $\times$ denote Cartesian products of sets. 
Finally, the union of finitely many sets $A_i$, indexed by a finite set $S$, is denoted by $\cup_{i \in S} A_i$.

%For any set $S$, we use $\Delta(S)$ to denote the set of all probability measure over $S$. For any real number $x$, by $|x|$ we denote its absolute value, while for a set $S$, $|S|$ denotes its cardinality. For any probability measure $\mu$ over a discrete set $X$,  we use $L^1(\mu)$ to denote all functions that satisfies $\sum_{x\in X} |f(x)|\mu(x)<\infty$. Similarly, $\{0\}_{i\in S}$ represents a $|S|$-dimensional zero vector. For a function $G:S\rightarrow \mathds{R}$, by $\{G(i)\}_{i\in S}$ we represent a vector of dimension $|S|$. For two real numbers $r_1,r_2$, $[r_1,r_2]$ represents the standard closed interval. For two integers $k_1, \ k_2$, such that $k_1<k_2$, we use $[k_1:k_2]$ to denote the set of integers in the interval $[k_1,k_2]$. For a vector $h\in \mathds{R}^n$, the $1$-Euclidean norm is denoted by $\|h\|_1$. For an event $T$ of an experiment, we use $\mathds{P}[T]$ to denote the probability that the event $T$ occurs. Symbols $\prod$ and $\times$ denote the Cartesian product of sets. The union of finitely many sets $A_i$ indexed by a finite set $S$ is denoted by $\cup_{i\in S}A_i$.

\section{Background and Problem Formulation} \label{problem_formulation}
We consider a time-inhomogeneous Markov decision process (MDP) with a finite state space $\mathcal{X}$ and a finite action space $\mathcal{A}$. The state space is partitioned into three disjoint subsets: the set of \emph{goal} or \emph{target} states $E$, the set of \emph{forbidden} or \emph{unsafe} states $U$, and the \emph{living} set $H := \mathcal{X} \setminus (E \cup U)$. The goal states are terminal, meaning that once the process enters the set, the evolution terminates. 

A system trajectory is an element $\omega = (x_0, a_0, x_1, a_1, \ldots)$ of the sample space $\Omega := (\mathcal{X} \times \mathcal{A})^{\infty}$, equipped with the $\sigma$-algebra $\mathcal{F}$ generated by the coordinate maps $X_t(\omega) = x_t$ and $A_t(\omega) = a_t$. Throughout, uppercase symbols $X_t$ and $A_t$ denote random variables, whereas lowercase symbols $x_t$ and $a_t$ denote their realizations. Suppose $\tau_S(t,x)$ denotes the first hitting time after $t$ of a set $S$ starting from state $X_t=x$.

\begin{ass}
    There exists an almost sure upper bound $T_{max}<\infty$ on the hitting time $\tau:=\tau_{E\cup U}(t,x)$ for all $(t,x)$.
\end{ass}

We consider a time-varying policy $\pi:=\{\pi_t\}_{t=0}^{T_{max}-1}$, where each $\pi_t$ assigns to every state $x\in\mathcal{X}$ a probability distribution over actions at time~$t$, i.e., $\pi_t \colon \mathcal{X} \rightarrow \Delta(\mathcal{A})$. For a state $x \in H$ and action $a \in \mathcal{A}$, the transition to the next state $y \in \mathcal{X}$ at time $t+1$ occurs with probability $P_{t,x,a}(y)$. The transition kernel induced by the policy $\pi$ is given by
\[
    P_{t,x}(y) := \sum_{a \in \mathcal{A}} \pi_t(a \mid x)\, P_{t,x,a}(y).
\]

For all $x\in H$ and $k\in [0:T_{max}-1]$, we define $\mathcal{P}_{k,x} := \{P_{k,x}(y)\}_{y \in \mathcal{X}}$ as the transition probability vector at time $t$ from state $x$ and $\mathcal{P}_k=[P_{k,x}(y)]_{x\in H, \ y\in \mathcal{X}}$ is the transition probability matrix at time $t=k$. We define $\mathscr{P}_t := \{\mathcal{P}_{k}\}_{k\in [t:T_{max}-1]}$ is a set of transition probability matrices from time $t$ until $T_{max}-1$.
%We define $\mathscr{P}_t := \{ {P}_{k,x}(y)\}_{y\in \mathcal{X}, \ x\in H, \ k\in [t:T_{max}]}$, where $\mathcal{P}_{k,x}:= \{P_{k,x}(y)\}_{y \in \mathcal{X}}$ for $k\in [0:\tau-1]$, and for $k\in [\tau: T_{max}]$, $\mathcal{P}_{k,x}=\{0\}_{y\in H}$. \raf{Why it is sometime $H$ and sometimes $X$} \abhijit{Because, the originating state is always assumed to be restricted to $H$ as from other states we are not interested in computing the safety function.}

If the transition probability $\mathscr{P}_t$ is precisely known, then probabilistic safety can be characterized by defining the \textit{safety function} as given in \cite{wisniewski2023probabilistic}.
\begin{defn}[Safety function] 
    Consider a fixed policy $\pi$ and a fixed transition probability $\mathscr{P}_t$. The safety function $S(x,t)$ for a state $X_t=x\in H$ is defined as the probability that any realization starting from a state $X_t=x$ hits the unsafe set $U$ before hitting the goal set $E$ following policy $\pi$, i.e.,
    \begin{equation*}
        S(t,x) := \mathds{P}_{\pi}[\tau_U(t,x) < \tau_E(t,x); \ {\mathscr{P}}_t\big|  X_t=x].
    \end{equation*}
%    where $\tau_S(x,\mathscr{P}_t)$ is the first hitting time of a set $S$ starting from state $X_t=x$ with a transition probability $\mathscr{P}_t$. 
\end{defn} 
For notational convenience, we write $\mathds{P}_{\pi}[\cdot; \mathscr{P}_t]$ for the probability measure over system trajectories induced by the policy $\pi$ and transition kernel $\mathscr{P}_t$, and $\mathds{E}_{\pi}[\cdot ; \mathscr{P}_t]$ for its associated expectation.

In this work, we assume that the exact transition probabilities are not known and it can vary over time and run. Instead, each $P_{t,x,a}(y)$ is assumed to lie within an interval or an ambiguity set, leading to an \emph{interval MDP} representation. Specifically, the transition probability  lies within 
the ambiguity set \([P^-_{t,x,a}(y),\, P^+_{t,x,a}(y)]\), and that the width of %\raf{to complex notation try to simplify; why $\mathscr{P}_t^{\mathrm{int}}$ is not $y$-dependent }
this ambiguity set is bounded by \(\delta\), i.e.,
\[
    P^-_{t,x,a}(y) \le P_{t,x,a}(y) \le P^+_{t,x,a}(y),
    \qquad
    P^+_{t,x,a}(y) - P^-_{t,x,a}(y) \le \delta.
\]
 %The corresponding interval-valued transitions under a policy $\pi$ induce an interval kernel 
 %$\mathscr{P}_t^{\mathrm{int}}$, which replaces the precise kernel $\mathscr{P}_t$ in the subsequent safety analysis.

%\raf{This will be again an interval}
%\raf{Do you want this time inhomogeneous complication. What do you mean conditioned over probability measure $\mathscr{P}_t$}
Since we fix the policy, effectively, we have a Markov chain. Thus, in the rest of the paper, we consider a Markov chain with ambiguous transition probabilities. To construct the ambiguity set for the transition probability matrix $\mathscr{P}_t$, we define the $1$-Wasserstein distance. For any two distribution $\mu, \nu \in \Delta(|\mathcal{X}|)$ defined $W(\mu, \nu)$ as given below:
     \begin{equation*}
    \begin{split}
    & W(\mu, \nu)  :=  \Big{\{} \underset{\Gamma\in \Delta(\mathcal{X} \times \mathcal{X})}{\text{min }} \ \sum_{(y,z)\in \mathcal{X} \times \mathcal{X}} \Gamma(y,z) d(y,z)  \\
    &  \text{s.t.} \ \sum_{z\in \mathcal{X}} \Gamma(y,z)= \mu(y), \ \sum_{y\in \mathcal{X}} \Gamma(y,z)= \nu(z) \Big{\}},
\end{split}
\end{equation*}
%\mnl{
%\[
%W\big(\tilde{\mathcal{P}}_{t,x},\mathcal{P}_{t,x}\big)
%:= \inf_{\Gamma\in\Pi\big(\tilde{\mathcal{P}}_{t,x},\mathcal{P}_{t,x}\big)}
%\sum_{y,z\in\mathcal{X}} \Gamma(y,z)\, d(y,z)
%\]
%subject to
%\[
%\sum_{z\in\mathcal{X}} \Gamma(y,z)=\tilde{\mathcal{P}}_{t,x}(y)
%\quad\text{for all }y\in\mathcal{X},\qquad
%\sum_{y\in\mathcal{X}} \Gamma(y,z)=\mathcal{P}_{t,x}(z)
%\quad\text{for all }z\in\mathcal{X},
%\]
%where \(\Pi(\tilde{\mathcal{P}}_{t,x},\mathcal{P}_{t,x})\) is the set of couplings (transport plans) on \(\mathcal{X}\times\mathcal{X}\) with the prescribed marginals.
%}
%\raf{Are you sure that you want $\Gamma(y,z)= {P}_{t,x}$, one is a real number, the second one is the vector}\abhijit{But both $\Gamma(y,z)$ and $\tilde{P}_{t,x}(y)$ are real.}
where $d(x,y)$ is a distance metric. In this work, we consider $d(x,y)$ to be the \emph{Hamming distance} between $x,y \in \mathcal{X}$ defined as
\[
d(x,y) := 
\begin{cases}
0, & x = y, \\
1, & x \neq y.
\end{cases}
\]
Considering $d(x,y)$ as the Hamming distance, the $1$-Wasserstein distance $W(\mu,\nu)$ becomes equivalent to the \textit{total variation distance distance} $TV(\mu,\nu)$. The total variation distance $TV(.)$ plays a crucial role in obtaining our final result.

The Wasserstein distance can be represented in a dual form using Kantorovich's duality as follows \citep{gao2023distributionally}: 
\small
\begin{equation*}
\begin{split}
&W(\mu,\nu)\\
& = \underset{f\in {L}^1(\mu), g\in {L}^1(\nu)}{\text{sup}}
   \begin{cases}
    %\sum_{z_1\in \mathcal{X}} f(z_1) \tilde{{P}}_{x,a}(z_1) + \sum_{z_2\in \mathcal{X}} g(z_2) {P}_{x,a}(z_2), \\
     \langle f,\mu \rangle + \langle g, \nu \rangle  \\
    s.t. \ f(z_1) + g(z_2) \leq d(z_1, z_2), \ \forall z_1,z_2 \in \mathcal{X}
    %s.t. \ f(z_1) + g(z_2) \leq |z_1-z_2|, \ \forall z_1,z_2 \in \mathcal{X}
    \end{cases} 
  \end{split}
\end{equation*}
\normalsize
where $ \langle f,\mu \rangle := \sum_{y\in \mathcal{X}} f(y)\mu(y)$. \\
\\
Then, we get the following expression for the Wasserstein distance $W(\mu,\nu)$:
\small
\begin{equation}
\begin{split}
& W(\mu,\nu)\\
    & = \underset{f\in {L}^1(\mu), g\in {L}^1(\nu)}{\text{sup}}
   \begin{cases}
   \langle f,\mu \rangle + \langle g,\nu \rangle \\
    s.t. \  g(z_2) \leq \underset{z_1}{\text{inf}} \ \big( d(z_1, z_2) - f(z_1) \big) , \ \forall z_2 \in \mathcal{X} 
    \end{cases} \\
    & = \underset{f\in {L}^1(\mu)}{\text{sup}} \ 
    \langle f,\mu \rangle +  \sum_{z_2\in \mathcal{X}} \underset{z_1}{\text{inf}} \ \big( d(z_1, z_2) - f(z_1) \big) {P}_{t,x}(z_2) 
  \end{split}
  \label{Wasserstein_dual2}
\end{equation}
\normalsize

%\mnl{Kantorovich duality via Lipschitz functions}
%\begin{align}
     %W(\tilde{\mathcal{P}}_{x},\mathcal{P}_{x})=\sup_{||f||_{Lip}\leq 1}\langle \tilde{\mathcal{P}}_{x}-\mathcal{P}_{x},f \rangle
%\end{align}
%with the Lipschitz seminorm defined as $||f||_{Lip}=\max_{z \neq y} \frac{|f(z)-f(y)|}{d(z,y)}$.

Our main objective is to derive an upper bound on the safety function for the IMDP using only a finite number of data samples. 
Before doing so, we first assume access to the true nominal transition probability matrix $\mathscr{P}_t$ together with a Wasserstein-type ambiguity set $\mathcal{D}^{\delta}(\mathscr{P}_{t})$. 
This ambiguity set is defined under the following rectangularity assumption \citep{iyengar2005robust, liu2022distributionally}. 
The rectangularity assumption is essential for deriving the recursive expression of the \textit{robust safety function} introduced later.

    We assume that the ambiguity set $\mathcal{D}^{\delta}(\mathcal{P}_{k})$ is defined as follows: 
 \begin{equation*}
\begin{split}
     & \mathcal{D}^{\delta}(\mathcal{P}_{k}) := \{\tilde{\mathcal{P}}_k \mid \underset{x\in \mathcal{X}}{\sup} \ W\left(\tilde{\mathcal{P}}_{k,x},\mathcal{P}_{k,x} \right)\leq \delta\}, \\
    %& \mathscr{D}^{\delta}(\mathscr{P}_{t}) := \underset{(k,x)\in [t:T_{max}-1]\times H}{\prod} \mathcal{D}^{\delta}(\mathcal{P}_{k,x}), \\
%    & \mathscr{D}^{\delta}(\mathscr{P}_{t}) := \Big{\{} \tilde{\mathcal{P}}_{k,x} \Big| \tilde{\mathcal{P}}_{k,x}\in \mathcal{D}^{\delta}(\mathcal{P}_{k,x}) \Big{\}}_{(k,x)\in [t:T_{max}-1]\times H}, \\
   %& \text{where} \ \forall (k,x)\in [0,\tau-1]\times H, \\
   % & \mathcal{D}^{\delta}(\mathcal{P}_{k,x}) := \Big{\{}\tilde{\mathcal{P}}_{k,x}\in \Delta(\mathcal{X}) \ \big| \ W(\tilde{\mathcal{P}}_{k,x},\mathcal{P}_{k,x}) \leq \delta  \Big{\}}, \\
  %  & \text{and} \ \forall (k,x)\in [\tau,T_{max}]\times H, \  \mathcal{D}^{\delta}(\mathcal{P}_{k,x}) = \{0\}_{y\in \mathcal{X}}.
    \end{split}
\end{equation*}
\begin{ass}[Rectangularity assumption]
For the family of transition probability matrices from time $t$ to $T_{max}-1$, we define the following ambiguity set:
\begin{equation*}
    \begin{split}
       & \mathcal{D}^{\delta}(\mathscr{P}_{t}) := \cup_{k\in [t:T_{max}-1]} \ \mathcal{D}^{\delta}\left(\mathcal{P}_k \right) \\  
    \end{split}
\end{equation*}
 
 Similarly, for a distribution $\mathcal{P}_{k,x}$, $(k,x)\in [t:T_{max}-1]\times H$, we define the following ambiguity set:
\begin{equation*}
\mathcal{D}^{\delta}(\mathcal{P}_{k,x}) := \Big{\{}\tilde{\mathcal{P}}_{k,x}\in \Delta(\mathcal{X}) \ \big| \ W(\tilde{\mathcal{P}}_{k,x},\mathcal{P}_{k,x}) \leq \delta  \Big{\}}.
\end{equation*}
%Thus, for a given $X_t=x$, $\mathscr{D}^{\delta}(\mathscr{P}_{t})$ can be expressed as follows:
%\begin{equation}\label{rectan_prop}
%\begin{split}
%   \mathscr{D}^{\delta}(\mathscr{P}_{t})& = \mathcal{D}^{\delta}(\mathcal{P}_{t,x}) \times \mathscr{D}^{\delta}(\mathcal{P}_{t+1}). 
%\end{split}
%\end{equation}
\end{ass}
%where $\forall (k,x)\in [0:\tau-1]\times H,$
%\begin{equation*}
%\mathcal{D}^{\delta}(\mathcal{P}_{k,x}) := \Big{\{}\tilde{\mathcal{P}}_{k,x}\in \Delta(\mathcal{X}) \ \big| \ W(\tilde{\mathcal{P}}_{k,x},\mathcal{P}_{k,x}) \leq \delta  \Big{\}},
%\end{equation*}
%and $\forall (k,x)\in [\tau:T_{max}-1]\times H,$
%\begin{equation*}
 %  \mathcal{D}^{\delta}(\mathcal{P}_{k,x}) = \{0\}_{y\in \mathcal{X}}. 
%\end{equation*}
%\abhijit{I think, set $\mathcal{D}^{\delta}(\mathcal{P}_{k,x})$ has to be define for all $k\in [0,T_{max}-1]$ as follows:
%where 
%\begin{equation*}
%\mathcal{D}^{\delta}(\mathcal{P}_{k,x}) := \Big{\{}\tilde{\mathcal{P}}_{k,x}\in \Delta(\mathcal{X}) \ \big| \ W(\tilde{\mathcal{P}}_{k,x},\mathcal{P}_{k,x}) \leq \delta  \Big{\}},
%\end{equation*}
%}\\
%\abhijit{Write about the rectangularity assumption}
\par To characterize safety with ambiguous transition probability, we now introduce the distributionally robust safety function. 
\begin{defn}[Robust safety function] \label{robust_safety}
    At time index $t$, consider a fixed policy $\pi$ and a given nominal transition probability matrix $\mathscr{P}_t$ with an ambiguity set $\mathcal{D}^{\delta}(\mathscr{P}_t)$. For a state $X_t=x$, we define the \text{distributionally robust safety function} or simply \text{robust safety function} $\mathcal{S}^{R}(t,x)$ as the worst-case probability that any realization starting from $X_t=x\in H$ visits the unsafe set $U$ before visiting the goal set $E$, i.e.,
    \begin{equation}
       % \mathcal{S}^{\mathscr{P},\delta}_{\pi}(x)
        \mathcal{S}^{R}(t,x):= \underset{\tilde{\mathscr{P}_t}\in \mathcal{D}^{\delta}(\mathscr{P}_t)}{\text{sup}} \ \mathds{P}_{\pi} \left[\tau_U(t,x) < \tau_E(t,x) \ ; \ \tilde{\mathscr{P}}_t \big| X_t=x\right] 
        \label{dr_safety_fn}
    \end{equation}
\end{defn}
%\mnl{I would suggest a simple notation ${S}_{\pi}(x;\delta)$}
\par We consider the following notion of distributionally robust probabilistic safety, namely robust $p$-safety. This definition is inspired by the $p$-safety notion for standard MDPs introduced in \cite{wisniewski2023probabilistic}.  
\begin{defn}[Robust $p$-safety] 
    Suppose the policy $\pi$, robustness parameter $\delta$ and the safety parameter $p\in (0,1)$ are fixed. We call an MDP to be robust $p$-safe if $\underset{t\in [ 0:T_{max}-1],x\in H}{{\max}}\ \mathcal{S}^{R}(t,x)\leq p$. 
\end{defn}
\par Our goal is to assess the robust $p$-safety of the IMDP for a given \textit{evaluation policy} $\pi$ by computing the robust safety function $S^R(t,x)$. 
The nominal transition probability matrix $\mathscr{P}$, the robustness parameter $\delta$, and the safety threshold $p$ are assumed to be known. 
The robust safety function, as defined in \eqref{dr_safety_fn}, is generally difficult to compute because it requires optimization over probability distributions. 
Our objective is to make this computation tractable by reformulating the problem as a finite-dimensional dual optimization problem. 
To achieve this, we employ the Kantorovich duality for the Wasserstein distance.

\section{Distributionally robust safety verification}\label{DR safety verification}
In this section, we present derive various useful expressions for the robust safety function to evaluate robust $p$-safety.

 \begin{lem}[Finite-horizon expression]
 \label{dr_safety_fn_rew}
  For a state $X_t=x$ and time step $t\in [0:T_{max}-1]$, the robust safety function $\mathcal{S}^{R}(t,x)$ can be expressed as:
  \begin{equation}\label{rob_safety_fn_2}
     \mathcal{S}^{R}(t,x) = \underset{\tilde{\mathscr{P}}_t\in \mathcal{D}^{\delta}(\mathscr{P}_t)}{\text{sup}} \ \mathds{E}_{\pi} \Big[ \sum_{k=t}^{\tau(t,x)+t-1} \kappa(X_k,\tilde{\mathcal{P}}_{k,X_k}) \ ; \ \tilde{\mathscr{P}}_t \big| X_t=x \Big],
  \end{equation}
  %\abhijit{I think $\mathscr{D}^{\delta}(\mathscr{P}_t)$ has to be a deterministic set.}\\
  where,
  \begin{equation*}
      \begin{split}
          \kappa(x,\tilde{\mathcal{P}}_{k,x}):= \sum_{y\in U} \tilde{P}_{k,x}(y).
         % \kappa(k,x):= \sum_{y\in U} \tilde{P}_{k,x}(y).
          %\ \text{and} \ \tau(x,\tilde{\mathscr{P}}_t):= \tau_{U\cup E}(x,\tilde{\mathscr{P}}_t).
          %\begin{cases}
          %    1, \text{ if } X_{t+1}\in U \\
           %   0, \text{ otherwise. }
          %\end{cases}
      \end{split}
  \end{equation*}
 \end{lem}
 %\mnl{Remark that $c_{t+1}=I_U(X_{t+1})$.}
 \begin{pf}
    Using Corollary $1$ in \cite{wisniewski2023probabilistic} and Definition \ref{robust_safety}, we get the desired result. \QEDA
 \end{pf}

 \par We now show that the robust safety function $S^{R}_{\pi}$ can also be computed recursively using dynamic programming.
 \begin{lem}[Recursive expression]
 \label{dr_q_dp}
     The robust safety function $S^{R}_{\pi}(t,x)$ is given by
     %\small
  \begin{equation*}
  \begin{split}
     S^{R}(t,x)  =& \underset{\tilde{\mathcal{P}}_{t,x}\in \mathcal{D}^{\delta}(\mathcal{P}_{t,x})}{\text{sup}}  \mathds{E}_{\pi} \Big[\kappa(x,\tilde{\mathcal{P}}_{t,x}) +  S^{R}_{\pi}(t+1,y) \ ; \ \tilde{\mathscr{P}}_t \big|   y\sim \tilde{\mathcal{P}}_{t,x} \Big].
     \end{split}
  \end{equation*}
  \normalsize
 %where, $\kappa(t,x)=\sum_{y\in U} \tilde{P}_{t,x}(y)$. 
 \end{lem}
\begin{pf}
 From Lemma \ref{dr_safety_fn_rew}, 
 \small
 \begin{equation*}
     \begin{split}
         & \mathcal{S}^{R}(t,x)\\
         & = \underset{\tilde{\mathscr{P}}_t\in \mathcal{D}^{\delta}(\mathscr{P}_t)}{\text{sup}} \ \mathds{E}_{\pi} \Big[ \sum_{k=t}^{\tau(t,x)+t-1} \kappa(X_k,\tilde{\mathcal{P}}_{k,X_k}) \ ; \ \tilde{\mathscr{P}}_t \big| X_t=x \Big]\\
         & \overset{(a)}{=} \underset{\tilde{\mathscr{P}}_t\in \mathcal{D}^{\delta}(\mathscr{P}_t)}{\text{sup}} \ \mathds{E}_{\pi} \Big[ \kappa(x,\tilde{\mathcal{P}}_{t,x}) + \sum_{k=t+1}^{\tau(t+1,X_{t+1})+(t+1)-1} \kappa(X_k,\tilde{\mathcal{P}}_{k,X_k})  ;  \tilde{\mathscr{P}}_t \big| X_t=x \Big]\\
         & \overset{(b)}{=} \underset{\tilde{\mathcal{P}}_{t,x}\in \mathcal{D}^{\delta}(\mathcal{P}_{t,x})}{\text{sup}} \ \mathds{E}_{\pi} \Big[ \kappa(x,\tilde{\mathcal{P}}_{t,x}) + \\
         & \underset{\tilde{\mathscr{P}}_{t+1}\in \mathcal{D}^{\delta}(\mathscr{P}_{t+1})}{\text{sup}}  \mathds{E}_{\pi} \big[\sum_{k=t+1}^{\tau(t+1,X_{t+1})+(t+1)-1} \kappa(X_k,\tilde{\mathcal{P}}_{k,X_k})  ;  \tilde{\mathscr{P}}_t \big| X_{t+1} \big]  ;  \tilde{\mathscr{P}}_t \ \big| X_t=x \Big] \\
         & = \underset{\tilde{\mathcal{P}}_{t,x}\in \mathcal{D}^{\delta}(\mathcal{P}_{t,x})}{\text{sup}}  \mathds{E}_{\pi} \Big[\kappa(x,\tilde{\mathcal{P}}_{t,x}) +  S^{R}_{\pi}(t+1,y) \ ; \ \tilde{\mathscr{P}}_t  \big|  y\sim \tilde{\mathcal{P}}_{t,x} \Big]
     \end{split}
 \end{equation*}
 \normalsize
 The inner expectation in relation $(a)$ is in view of Lemma $2$ in \cite{mazumdar2024safe}. Relation $(b)$ holds in view of the rectangularity structure of the ambiguity set $\mathscr{D}^{\delta}(\mathscr{P}_t)$.  \QEDA
\end{pf}
We will also use the following expression for the robust safety function.
\begin{lem}
    The safety function $S^{R}(t,x)$ can be expressed as follows:
    \small
    \begin{equation*}
        S^{R}(t,x) = \underset{\tilde{\mathcal{P}}_{t,x}\in \mathcal{D}^{\delta}(\mathcal{P}_{t,x})}{\text{sup}}  \mathds{E}_{\pi} \Big[ c(x,y) + S^{R}(t+1,y)  ;  \tilde{\mathscr{P}}_t \big| X_t=x, y\sim \tilde{\mathcal{P}}_{t,x} \Big],
    \end{equation*}
    \normalsize
    where, $c(x,y)=\begin{cases}
        & 1, \ \text{if } y\in U \\
        & 0, \ \text{otherwise}.
    \end{cases}$
\end{lem}
\begin{pf}
    We observe that \[\mathds{E}_{\pi} \left[c(x,y) \ ; \ \tilde{\mathscr{P}}_t \big|  X_t=x \right]=\mathds{E}_{\pi} \left[\kappa(x,\tilde{\mathcal{P}}_{t,x}) \ ; \ \tilde{\mathscr{P}}_t \big| X_t=x \right].\]
    Thus, from Lemma \ref{dr_q_dp}, we get the desired result. \QEDA
\end{pf}
\subsection{Convex program formulation of the robust safety function:}
In the following lemma, we express the robust safety function as a finite-dimensional optimization problem. To this end, we make use of the Kantorovich duality result for Wasserstein distance computation.
\begin{lem}[Dual form of the robust safety function]
\label{Robust_safety_dual}
    The distributionally robust safety function is the solution of:
    \small
    \begin{equation}
    \begin{split}
       & S^{R}(t,x)= \underset{\lambda \geq 0}{\text{inf}} \ \Big[\lambda \delta   + \sum_{y\in \mathcal{X}} \ \underset{l\in \mathcal{X}}\max \ \Big( -\lambda d(l,y) + c(x,l) + S^{R}(t+1,l) \Big) {P}_{t,x}(y) \Big].
    \label{robsut_Q}
    \end{split}
\end{equation}
\normalsize
\end{lem}

\begin{pf}
    From the definition, $S^{R}_{\pi}(x)$ is the solution of the above optimization problem
    %\small
    \begin{equation}
        \begin{split}
           & S^{R}(t,x) \\
           =  & \underset{{\tilde{\mathcal{P}}_{t,x}}\in \mathcal{D}^{\delta}(\mathcal{P}_{t,x})}{\text{sup}} \mathds{E}_{\pi} \big[ c(x,y) + S^{R}(t+1,y) \ ; \ \tilde{\mathcal{P}}_{t,x} \big|  y\sim \tilde{\mathcal{P}}_{t,x} \big] \\
             & = \begin{cases}
                    \underset{{\tilde{\mathcal{P}}_{t,x}}\in \Delta(\mathcal{X})}{\text{sup}} \mathds{E}_{\pi} \big[ c(x,y) + S^{R}(t+1,y) \ ; \ \tilde{\mathcal{P}}_{t,x} \big|  y\sim \tilde{\mathcal{P}}_{t,x} \big] \\
                    \text{s.t. } W(\tilde{\mathcal{P}}_{t,x},\mathcal{P}_{t,x}) \leq \delta
                \end{cases}  \\
                & \overset{(a)}{=} \begin{cases}
                    \underset{{\tilde{{P}}_{t,x}}\in \Delta(\mathcal{X})}{\text{sup}} \mathds{E}_{\pi} \big[ c(x,y) +  S^{R}(t+1,y) \ ; \ \tilde{\mathcal{P}}_{t,x} \big|  y\sim \tilde{\mathcal{P}}_{t,x} \big] \\
                    \text{s.t. }   \sum_{z_1\in \mathcal{X}} f(z_1) \tilde{{P}}_{t,x}(z_1) \\
                    + \sum_{z_2\in \mathcal{X}} \underset{z_1}{\text{inf}} \ \big( d(z_1,z_2) - f(z_1) \big) {P}_{t,x}(z_2) \leq \delta; \ \forall f \in L^1
                \end{cases},  
        \end{split}
    \end{equation}
    \normalsize
    %where $Z^{x}_t:=\{\tilde{\mathcal{P}}_{t,x}, y\sim \tilde{\mathcal{P}}_{t,x}$\}. 
    %\mnl{This notation is strange. In my view if $y\sim \tilde{\mathcal{P}}_{t,x}$ then of course $X_t=x$}\\
    %\abhijit{Thats true. Can you suggest another notation?}\\
In the above expression, equality $(a)$ is due to \eqref{Wasserstein_dual2}. Using the standard duality in constrained optimization, we further get $\forall f \in L^1$:
\small
    \begin{equation}
        \begin{split}
           & S^{R}(t,x) \\
         & = \underset{\lambda \geq 0}{\text{inf}} \   \underset{{\tilde{\mathcal{P}}_{t,x}}\in \Delta(\mathcal{X})}{\text{sup }}  \sum_{y\in \mathcal{X}} \left[  c(x,y) +  S^{R}(t+1,y) \right] \tilde{{P}}_{t,x}(y) \\
                 &   - \lambda \Big(  \sum_{z_1\in \mathcal{X}} f(z_1) \tilde{{P}}_{t,x}(z_1) + \sum_{z_2\in \mathcal{X}} \underset{z_1}\min \ \big( d(z_1,z_2) - f(z_1) \big) {P}_{t,x}(z_2) - \delta \Big)\\
                % & \ \forall f \in L^1 \\
        & = \underset{\lambda \geq 0}{\text{inf}} \   \underset{{\tilde{\mathcal{P}}_{t,x}}\in \Delta(\mathcal{X})}{\text{sup }}  \sum_{y\in \mathcal{X}} \left[ c(x,y) +  S^{R}(t+1,y) \right] \tilde{{P}}_{t,x}(y)  \\
        & - \lambda \Big(  \sum_{z_1\in \mathcal{X}} f(z_1)  \tilde{{P}}_{t,x}(z_1)  + \sum_{z_2\in \mathcal{X}} \underset{z_1}\min \ \big( d(z_1,z_2) - f(z_1) \big)  {P}_{t,x}(z_2) - \delta \Big) \\
        %& \ \forall f \in L^1 \\
        & = \underset{\lambda \geq 0}{\text{inf}}   \underset{{\tilde{\mathcal{P}}_{t,x}}\in \Delta(\mathcal{X})}{\text{sup }}  \sum_{y\in \mathcal{X}} \big[  c(x,y)  +  S^{R}(t+1,y)  - \lambda f(y)   \big] \tilde{{P}}_{t,x}(y) \\
                 &   + \lambda \Big(  \sum_{z_2\in \mathcal{X}} \underset{z_1}\max \ \big( -d(z_1,z_2) + f(z_1) \big)  {P}_{t,x}(z_2) + \delta \Big) \\ %\forall f \in L^1 \\
      &  \overset{a}{=} \underset{\lambda \geq 0}{\text{inf}} \ \Big[\lambda \delta  + \sum_{y\in \mathcal{X}} \ \ \underset{l\in \mathcal{X}}\max \ \Big( -\lambda d(l,y) + c(x,l) +  S^{R}(t+1,l) \Big) {P}_{t,x}(y) \Big]
        \end{split}
    \end{equation}
    \normalsize
To get equality $(a)$ in the above expression, we use the following analysis: since the expression is valid for all $f\in L^1$, we consider $f(l)$ such that $\lambda f(l)=-(c(x,l) + S^{R}_{\pi}(t+1,l))$ for all $y\in \mathcal{X}$. This choice of $f(y)$ helps us to avoid $\tilde{P}_{t,x}(y)$, thus eliminating the inner optimization. 
\end{pf}
\section{Data-driven robust safety function}
\vspace{-0.2cm}
%In this section, we present a data-driven approach to compute an upper bound on the robust safety function for the interval-MDP. Suppose that we do not have access to $\mathcal{P}_{t,x}$ for each state $x\in H$ and $t\in [0:T_{max}-1]$. What we have is data samples of length $N$ given by  $\{{\xi}_{t,x,1}, {\xi}_{t,x,2}, ..., {\xi}_{t,x,N}\}$ which are got by running the system $N$ times \raf{a bit clumsy}. 
 %But each time we run the system, for each $t$ and $x$, the true model $\mathcal{P}_{t,x}$ changes i.e., each data point $\{\xi_{t,x,i}\}$ is generated by a transition probability $\mathcal{P}_{t,x,i}$ such that $W(\mathcal{P}_{t,x,i},\mathcal{P}_{t,x,j})\leq \delta$ for all $i,j\leq N$. To this end, we define the time-average empirical mean as 
 %For each $t$ and $x$, each data point $\{\xi_{t,x,i}\}$ is generated by a transition probability $\mathcal{P}_{t,x,i}$ such that $W(\mathcal{P}_{t,x,i},\mathcal{P}_{t,x,j})\leq \delta$ for all $i,j\leq N$. Thus, we can no longer use standard results from concentration inequality that relate the empirical distribution calculated from i.i.d. data samples to the true distribution. To this end, we define the time-average empirical mean as 
  %  \[
  % \hat{{P}}_{t,x}(y)= \frac{1}{N} \sum_{i=1}^N \mathds{1}\{\xi_{t,x,i}=y\}. 
  %  \]
   % Also define the true average distribution $\bar{P}_{t,x}(y)$ as
   % \[
% \bar{P}_{t,x}(y) := \frac{1}{N} \sum_{i=1}^{N} P_{t,x,i}(y)
  %  \]
%\abhijit{How do we modify Lemma 3 replacing $P_{t,x}(y)$ by $\hat{P}_{t,x}(y)$ such that $S^{DR}_{t,x}$ can be upper bounded for each $t$  and $x$}\\
In this section, we present a data-driven approach to compute an upper bound on the robust safety function for the interval MDP. We assume that the true transition probabilities ${P}_{t,x}(.)$ are not directly accessible for each state $x\in H$ and time $t\in [0:T_{\max}-1]$. Instead, for every pair $(t,x)$ we observe $N$ successor states
\[
    \{\xi_{t,x,1}, \xi_{t,x,2}, \dots, \xi_{t,x,N}\},
\]
obtained by running the system $N$ times. For each $i\in\{1,\dots,N\}$, the sample $\xi_{t,x,i}$ is generated according to a transition probability vector $\mathcal{P}_{t,x,i} := \{P_{t,x,i}(y)\}_{y\in\mathcal{X}}$, and we assume that these distributions are pairwise close in Wasserstein distance, i.e.,
\[
    W\big(\mathcal{P}_{t,x,i},\mathcal{P}_{t,x,j}\big)\le \delta
    \qquad \text{for all } i,j\in\{1,\dots,N\}.
\]
Thus, the samples are independent but not identically distributed, and we cannot directly apply standard concentration inequalities for i.i.d.\ data that relate an empirical distribution to a single underlying “true’’ distribution.

To proceed, we define the empirical averaged distribution at $(t,x)$ as
\[
    \hat{P}_{t,x}(y)
    := \frac{1}{N} \sum_{i=1}^N \mathds{1}\{\xi_{t,x,i} = y\},
    \qquad y\in\mathcal{X},
\]
and the corresponding (unknown) averaged true distribution
\[
    \bar{P}_{t,x}(y)
    := \frac{1}{N} \sum_{i=1}^N P_{t,x,i}(y),
    \qquad y\in\mathcal{X}.
\]

Suppose, $\forall (t,x)\in [0:T_{max}-1]\times H, \ \hat{\mathcal{P}}_{t,x} := \{\hat{P}_{t,x}(y)\}_{y\in \mathcal{X}}$ and $\hat{\mathcal{P}}_t:=[\hat{P}_{t,x}(y)]_{x\in H, y\in \mathcal{X}}$ be the empirical transition probability matrix at time step $t\in [0:T_{max}-1]$.
%For the rest of the paper, we define the following:
We define the following family of empirical state-transition matrices from time $k\in [t:T_{max}-1]$ as  $\hat{\mathscr{P}}_t:= \{\hat{\mathcal{P}}_{k}\}_{ k\in [t:T_{max}-1]}$.

Similar to ambiguity sets defined with the true nominal transition probabilities, empirical ambiguity sets are defined as follows. Ambiguity set around the empirical transition probability matrix is defined as:
\begin{equation*}
    \begin{split}
    & \mathcal{D}^{\delta+\rho}(\hat{\mathcal{P}}_{k}) := \{\tilde{\mathcal{P}}_k \mid \underset{x\in \mathcal{X}}{\sup} \ W\left(\tilde{\mathcal{P}}_{k,x},\hat{\mathcal{P}}_{k,x} \right)\leq \delta+\rho\}, \\
    \end{split}
\end{equation*}
where, $ \rho
    := \frac{|\mathcal{X}|}{2}\,\varepsilon, \
    \varepsilon
    := \sqrt{\frac{\ln(2|\mathcal{X}|/\beta)}{2N}},$ and $\beta\in (0,1)$ is the confidence parameter.

For the family of empirical transition probability matrices for $k\in [t:T_{max}-1]$, we define the following ambiguity set:
\begin{equation*}
    \begin{split}
       & \mathcal{D}^{\delta+\rho}(\hat{\mathscr{P}}_{t}) := \cup_{k\in [t:T_{max}-1]} \ \mathcal{D}^{\delta+\rho}\left(\hat{\mathcal{P}}_k \right) \\  
    \end{split}
\end{equation*}

Similarly, for an empirical average distribution $\hat{\mathcal{P}}_{k,x}$ $\forall (k,x)\in [t:T_{max}-1]\times H$ we define the following ambiguity set:
\begin{equation*}
\mathcal{D}^{\delta+\rho}(\hat{\mathcal{P}}_{k,x}) := \Big{\{}\tilde{\mathcal{P}}_{k,x}\in \Delta(\mathcal{X}) \ \big| \ W(\tilde{\mathcal{P}}_{k,x},\hat{\mathcal{P}}_{k,x}) \leq \delta + \rho  \Big{\}}.
\end{equation*}
%and $\forall (k,x)\in [\tau:T_{max}]\times H,$
%\begin{equation*}
 %  \hat{\mathcal{D}}^{\delta+\rho}(\hat{\mathcal{P}}_{k,x}) := \{0\}_{y\in \mathcal{X}}. 
%\end{equation*}
We now introduce the \textit{empirical robust safety function } $\hat{S}^{R}(t,x)$ for each $(t,x)\in [0:T_{max}-1]\times H$ as follows:
\begin{equation} \label{emp_rob_safety_fn}
 \hat{S}^{R}(t,x) := \underset{\tilde{\mathscr{P}}_t\in {\mathcal{D}}^{\delta+\rho}(\hat{\mathscr{P}}_t)}{\text{sup}} \ \mathds{E}_{\pi} \Big[ \sum_{k=t}^{\tau(t,x)+t-1} \kappa(X_k,\tilde{\mathcal{P}}_{k,X_k})  ;  \tilde{\mathscr{P}}_t \big| X_t=x \Big],
\end{equation}
recall that 
\begin{equation*}
      \begin{split}
          \kappa(x,\tilde{\mathcal{P}}_{k,x}):= \sum_{y\in U} \tilde{P}_{k,x}(y).
         % \kappa(k,x):= \sum_{y\in U} \tilde{P}_{k,x}(y).
          %\ \text{and} \ \tau(x,\tilde{\mathscr{P}}_t):= \tau_{U\cup E}(x,\tilde{\mathscr{P}}_t).
          %\begin{cases}
          %    1, \text{ if } X_{t+1}\in U \\
           %   0, \text{ otherwise. }
          %\end{cases}
      \end{split}
  \end{equation*}
Using Lemma \ref{Robust_safety_dual}, we have the following dual form for the data-driven robust safety function $\hat{S}^{R}(t,x)$.
\begin{lem}[Empirical robust safety function]
    For each $(t,x)\in [0:T_{max}-1]\times H$,
    \small
    \begin{equation*}
       \begin{split}
   & \hat{S}^{R}(t,x) \\ 
   & =
    \inf_{\lambda\ge 0}
    \Bigg\{
        \lambda\big(\delta+\rho\big)
        + \sum_{y\in\mathcal{X}}
        \Big[
            \max_{l\in\mathcal{X}}
            \big(-\lambda d(y,l) + c(x,l) + \hat{S}^{R}\left(t+1,y \right) \big)
        \Big]
        \hat{P}_{t,x}(y)
    \Bigg\}.
   \end{split} 
    \end{equation*}
    \normalsize
\end{lem}

As a corollary to the above result, we can further rewrite the robust safety function as a finite convex program, which is useful for computational aspects.
\begin{cor}[Convex-program formulation]
    We define epigraphical
auxiliary variables $h_y\in \mathds{R}, \ y\leq |\mathcal{X}|$. Then the  empirical robust safety function $\hat{S}^{R}(t,x)$ is the optimal value of the following convex program:
    %\small
    \begin{equation} \label{Conv_opt}
    \begin{split}    
       & \underset{\lambda \geq 0, \ h_y}{\text{inf}} \ \Big( \lambda (\delta + \rho) + \sum_{y\in \mathcal{X}} \ h_y   \Big) \\
      & \text{s.t.} \  \underset{l\in \mathcal{X}}\max \ \Big( -\lambda d(y,l) + c(x,l) + \hat{S}^{R}(t+1,l) \Big) \hat{P}_{t,x}(y) \leq h_y; \ \forall y \in \mathcal{X}. 
    \end{split}
    \end{equation}
    \normalsize
\end{cor}
We now present another important result, which provides a high-confidence upper bound on the true robust safety function for the original IMDP. 
\begin{thm}[Upper bound of the robust safety function]
%\label{lem:data_driven_upper_bound} 

Fix a time index $t\ge 0$, a state $x\in H$ and a confidence parameter $\beta\in (0,1)$. 
With probability at least $(1-\beta)$, %\raf{what is $\beta$}
\[
    S^{R}(t,x)
    \;\le\;
    \hat{S}^{R}(t,x).
\]
\normalsize
\end{thm}
\begin{pf}
    Using Hoeffding's inequality, for any $\varepsilon>0$ and for all $x\in H, \ t\in [0,T_{max}-1]$,
\[
    \mathbb{P}\Big(
        \big|\hat{P}_{t,x}(y)-\bar{P}_{t,x}(y)\big|
        \ge \varepsilon
    \Big)
    \;\le\;
    2\exp\big(-2N\varepsilon^2\big).
\]
%Take
%\[
%    \varepsilon_{N,\beta}
%    := \sqrt{\frac{\ln(2|\mathcal{X}|/\beta)}{2N}}.
%\]
Using the definition of $\varepsilon$,
\[
    2\exp\big(-2N\varepsilon^2\big)
    = 2\exp\Big(-\ln(2|\mathcal{X}|/\beta)\Big)
    = \frac{\beta}{|\mathcal{X}|}.
\]
By a union bound over all $y\in\mathcal{X}$,
\begin{equation*}
\begin{split}
&\mathbb{P}\Big(
    \max_{y\in\mathcal{X}}
    \big|\hat{P}_{t,x}(y)-\bar{P}_{t,x}(y)\big|
    \ge \varepsilon
\Big) \\
&
\;\le\;
\sum_{y\in\mathcal{X}}
\mathbb{P}\Big(
    \big|\hat{P}_{t,x}(y)-\bar{P}_{t,x}(y)\big|
    \ge \varepsilon
\Big) \\
&
\;\le\; \beta.
\end{split}
\end{equation*}
\normalsize
Thus, with probability at least $(1-\beta)$,
\[
    \max_{y\in\mathcal{X}}
    \big|\hat{P}_{t,x}(y)-\bar{P}_{t,x}(y)\big|
    \le \varepsilon.
\]
On this event,
\[
    \|\hat{\mathcal{P}}_{t,x}-\bar{\mathcal{P}}_{t,x}\|_1
    = \sum_{y\in\mathcal{X}}
        \big|\hat{P}_{t,x}(y)-\bar{P}_{t,x}(y)\big|
    \le |\mathcal{X}|\,\varepsilon.
\]

\begin{comment}
\emph{Step 2: From $\ell_1$ to Wasserstein with Hamming metric.}
\end{comment}
We can express the total variation distance by the following well-known result \citep{levin2017markov}:
\[
     \mathrm{TV}(\mu,\nu)
    = \frac{1}{2}\|\mu-\nu\|_1,
    \qquad \forall \mu,\nu\in\Delta(\mathcal{X}).
\]
Therefore, with probability at least $(1-\beta)$,
\begin{equation}
    W(\hat{\mathcal{P}}_{t,x},\bar{\mathcal{P}}_{t,x})
    =\mathrm{TV}(\hat{\mathcal{P}}_{t,x},\bar{\mathcal{P}}_{t,x}) = \frac{1}{2}\|\hat{\mathcal{P}}_{t,x}-\bar{\mathcal{P}}_{t,x}\|_1
    \le \frac{|\mathcal{X}|}{2}\,\varepsilon
    = \rho.
    \label{Wasser_emp_true}
\end{equation}

\begin{comment}
\emph{Step 3: Relating the true ambiguity set to a data-driven one.}
\end{comment}

By assumption the pairwise distances between the true models are bounded:
%\small
\[
    W(\mathcal{P}_{t,x,i},\mathcal{P}_{t,x,j}) \le \delta,
    \quad \forall i,j\leq N.
\]
\normalsize
Since the Wasserstein distance is convex in each argument, we can bound the distance between each $\mathcal{P}_{t,x,i}$ and the average model $\bar{\mathcal{P}}_{t,x}$ as follows:
%\small
\begin{equation} \label{Wasser_mean_indi}
\begin{split}
 W\big(\mathcal{P}_{t,x,i},\bar{\mathcal{P}}_{t,x}\big)  &  = W\Big(\mathcal{P}_{t,x,i}, \frac{1}{N}\sum_{j=1}^N \mathcal{P}_{t,x,j}\Big) \\
   & \le \frac{1}{N}\sum_{j=1}^N W\big(\mathcal{P}_{t,x,i},\mathcal{P}_{t,x,j}\big) 
    \le \delta.   
\end{split}
\end{equation}
%\[
%    W\big(\mathcal{P}_{t,x,i},\bar{P}_{t,x}\big)
%    = W\Big(\mathcal{P}_{t,x,i}, \frac{1}{N}\sum_{j=1}^N \mathcal{P}_{t,x,j}\Big)
%    \le \frac{1}{N}\sum_{j=1}^N W\big(\mathcal{P}_{t,x,i},\mathcal{P}_{t,x,j}\big)
%    \le \delta_0.
%\]
\normalsize
Thus each $\mathcal{P}_{t,x,i}$ lies in the ball of radius $\delta$ around $\bar{\mathcal{P}}_{t,x}$:
\[
    \mathcal{P}_{t,x,i}\in \mathcal{D}^{\delta}(\bar{\mathcal{P}}_{t,x}).
\]

%With probability at least $(1-\beta)$,
%\[
 %   W(\hat{P}_{t,x},\bar{P}_{t,x}) \le \rho_{N,\beta}.
%\]
By using relations \eqref{Wasser_emp_true},\eqref{Wasser_mean_indi} and the triangle inequality of $W$, we have that any $\tilde{\mathcal{P}}_{t,x}$ that satisfies
$W(\tilde{\mathcal{P}}_{t,x},\bar{\mathcal{P}}_{t,x})\le\delta$ also satisfies
\[
    W(\tilde{\mathcal{P}}_{t,x},\hat{\mathcal{P}}_{t,x})
    \le W(\tilde{\mathcal{P}}_{t,x},\bar{\mathcal{P}}_{t,x})
       + W(\bar{\mathcal{P}}_{t,x},\hat{\mathcal{P}}_{t,x})
    \le \delta + \rho.
\]
Therefore, $\forall (t,x)\in [0,T_{max}-1]\times H$,
\[
    \mathcal{D}^{\delta}(\bar{\mathcal{P}}_{t,x})
    \subseteq
    \hat{\mathcal{D}}^{\delta+\rho}(\hat{\mathcal{P}}_{t,x}).
\]

%The robust value functional is given by
%\[
 %   S^{R}(t,x)
 %   := \sup_{\tilde{\mathcal{P}}_{t,x}\in \mathcal{D}^{\delta}(\bar{\mathcal{P}}_{t,x})}
 %      \sum_{y\in\mathcal{X}} \left(c(x,y) + S^{R}(t+1,y) \right)\,\tilde{P}_{t,x}(y),
%\]

%By definition, the empirical robust safety function is given by,
%\[
 %   \hat{S}^{R}(t,x)
 %   = \sup_{\tilde{\mathcal{P}}_{t,x}\in \hat{\mathcal{D}}^{\delta+\rho}(\hat{\mathcal{P}}_{t,x})}
 %      \sum_{y\in\mathcal{X}} \left(c(x,y) + \hat{S}^{R}(t+1,y) \right)\,\tilde{P}_{t,x}(y).
%\]

Hence, with probability at least $(1-\beta)$,
\[
    \mathcal{D}^{\delta}({\mathscr{P}}_{t})
    \subseteq
    \hat{\mathcal{D}}^{\delta+\rho}(\hat{\mathscr{P}}_{t}), \ \forall t\in [0,T_{max}-1].
\]
Since taking the supremum over a larger set can only increase the value, from \eqref{rob_safety_fn_2} and \eqref{emp_rob_safety_fn},  we have the following result: with probability at least $1-\beta$,  
\[
    S^{R}(t,x)
    \;\le\;
    \hat{S}^{R}(t,x).
\]
\QEDA
\end{pf}
\begin{comment}
\begin{cor}
    The distributionally robust safety function can be expressed as follows:
    \small
    \begin{equation}
    \begin{split}
       & S^{DR}(x)= \underset{\lambda \geq 0}{\text{inf}} \ \Big[\lambda (\delta + \rho_{N,\beta})   + \sum_{y\in \mathcal{X}} \ \ \underset{l\in \mathcal{X}}\max \ \Big( -\lambda d(y,l) + c(x,l) +  S^{DR}(l) \Big) \hat{P}_{x}(y)  \Big]
    \label{robsut_Q}
    \end{split}
\end{equation}
\normalsize
\end{cor}
\end{comment}

%\vspace{-0.3cm}

\section{Numerical Example} \label{numerical results}

We consider a finite MDP with  
$\mathcal{X}=\{1,\dots,20\}$, partitioned into
$H=\{1,\dots,10\}$, $U=\{11,12\}$, $E=\{13,\dots,20\}$, where $U$ and $E$ are absorbing (terminal) sets. The action set is $\mathcal{A}=\{a_H,a_U,a_E\}$.

We consider that the process visits either the goal set $E$ or the unsafe set $U$ at time $t\leq T_{max}=10$ with probability $1$. For the ease of demonstration, we consider the nominal transition probabilities $P_{t,x,a}$ of the IMDP to be time invariant. Further, we consider a policy $\pi$ such that $\pi_t$ is same for all $t\leq T_{max}-2$ and for $\pi_{T_{max}-1}$ is chosen differently such that the process visits the target set $E$ or the unsafe set $U$ with at-most $T_{max}$ steps. For each $x\in H$, $a\in\mathcal{A}$, and $t\leq T_{max}-1$ the transition probabilities are: $P_{t,x,a_H}(y) =\frac{1}{|H|}\mathds{1}\{y\in H\}$,  $P_{t,x,a_U}(y)=\frac{1}{|U|}\mathds{1}\{y\in U\},$ and $P_{t,x,a_E}(y)=\frac{1}{|E|}\mathds{1}\{y\in E\}$. For $x\in U\cup E$, the process is absorbing:  
$P_{t,x,a}(y)=\mathds{1}\{y=x\}$ for all $a\in\mathcal{A}$ and $t\leq T_{max}$.

We evaluate the safety of the following randomized policy. For time steps $t\leq T_{max}-2$,
\[
\pi_t(a_H\mid x)=0.4,\ \
\pi_t(a_U\mid x)=0.3,\ \
\pi_t(a_E\mid x)=0.3,\ x\in H.
\]  
At the last step $t=T_{max}-1$, we use
\[
\pi_{T_{max}-1}(a_H\mid x)=0,\ \
\pi_{T_{max}-1}(a_U\mid x)=\pi_{T_{max}-1}(a_E\mid x)=0.5,
\]
which forces transitions from $H$ into $U\cup E$.

Thus, for the induced Markov chain satisfies, the nominal transition probabilities are as given below. For $x\in H$ and $t\leq T_{max}-2$,
\[
P_{t,x}(y)
=
\frac{0.4}{|H|}\mathds{1}\{y\in H\}
+
\frac{0.3}{|U|}\mathds{1}\{y\in U\}
+
\frac{0.3}{|E|}\mathds{1}\{y\in E\}.
\]
For $x\in H$ and $t=T_{max}-1$, the transition probabilities are:

\[
P_{t,x}(y)
=
\frac{0.5}{|U|}\mathds{1}\{y\in U\}
+
\frac{0.5}{|E|}\mathds{1}\{y\in E\}.
\]

%For the computation of the empirical robust safety function $\hat{S}^R(t,x)$, we assume that the empirical probabilities for the induced Markov chain is not accessible. Instead, we generate $100,000$ samples for each $\mathcal{P}_{x,a}$. We consider a confidence parameter of $\beta = 0.05$, ambiguity radius for the true transition probabilities $\delta=0.2$. With the above parameters, we first compute the empirical distribution $\mathcal{\hat{P}}_{x,a}$. Thereafter, we solve \eqref{Conv_opt} simultaneously for each $t\in [0:T_{max}-1]$ and $x\in H$ to get the empirical robust safety function. Following are the values of the empirical robust safety function $\hat{S}^R(t,x)$ for $4$-states:
For computing the empirical robust safety function $\hat{S}^R(t,x)$, 
we assume that the true empirical transition probabilities of the induced Markov chain are not directly available. 
Instead, for each $\mathcal{P}_{x,a}$, we generate $100{,}000$ transition samples. 
We use a confidence level of $\beta = 0.05$ and an ambiguity radius of $\delta = 0.2$ for the true transition probabilities. 
Based on these parameters, we first construct the empirical distributions $\hat{\mathcal{P}}_{x,a}$. 
Subsequently, we solve~\eqref{Conv_opt} independently for each $t \in [0 : T_{\max}-1]$ and $x \in H$ 
to obtain the empirical robust safety function. 
The corresponding values of $\hat{S}^R(t,x)$ for a $4$-state subset are listed below:

\begin{table}[h!]
\centering
\label{tab:robust_safety_values}
\[
\begin{array}{c|cccc}
\text{time } t & \hat{S}^R(t,1) & \hat{S}^R(t,2) & \hat{S}^R(t,3) & \hat{S}^R(t,4) \\
\hline
0 & 0.9305 & 0.9274 & 0.9309 & 0.9312 \\
1 & 0.9291 & 0.9286 & 0.9297 & 0.9278 \\
2 & 0.9291 & 0.9299 & 0.9296 & 0.9293 \\
3 & 0.9306 & 0.9288 & 0.9313 & 0.9292 \\
4 & 0.9296 & 0.9291 & 0.9259 & 0.9284 \\
5 & 0.9269 & 0.9259 & 0.9231 & 0.9247 \\
6 & 0.9187 & 0.9170 & 0.9200 & 0.9191 \\
7 & 0.9007 & 0.9012 & 0.9016 & 0.9018 \\
8 & 0.8608 & 0.8600 & 0.8638 & 0.8593 \\
9 & 0.7575 & 0.7607 & 0.7587 & 0.7553 \\
\end{array}
\]
\caption{Empirical robust safety function $\hat{S}^R(t,x)$.}
\end{table}

%The empirical distribution computed from the $100,000$ data samples are very close to the true nominal distribution.  Now, suppose we naively consider the empirical distribution without taking into account the ambiguity introduced by finite samples. That is, we treat the empirical distribution as the true distribution. We only consider the model ambiguity with radius $\delta=0.05$. Then the robust safety function is given below:

The empirical distributions obtained from the $100{,}000$ generated samples are nearly identical to the true nominal transition probabilities. 
Suppose, however, that we naively treat these empirical distributions as if they were the true model, thereby ignoring the statistical uncertainty arising from the finite number of samples. 
In this case, we account only for model ambiguity using a radius of $\delta = 0.2$. 
The resulting robust safety function is shown below:

\begin{table}[h!]
\centering
\label{tab:SR_values}
\[
\begin{array}{c|cccc}
\text{time } t & S^R(t,1) & S^R(t,2) & S^R(t,3) & S^R(t,4) \\
\hline
0 & 0.8342 & 0.8313 & 0.8343 & 0.8348 \\
1 & 0.8328 & 0.8321 & 0.8332 & 0.8317 \\
2 & 0.8327 & 0.8334 & 0.8334 & 0.8331 \\
3 & 0.8343 & 0.8326 & 0.8350 & 0.8330 \\
4 & 0.8336 & 0.8333 & 0.8302 & 0.8322 \\
5 & 0.8313 & 0.8305 & 0.8278 & 0.8293 \\
6 & 0.8250 & 0.8231 & 0.8261 & 0.8253 \\
7 & 0.8107 & 0.8111 & 0.8115 & 0.8115 \\
8 & 0.7798 & 0.7790 & 0.7828 & 0.7784 \\
9 & 0.6997 & 0.7029 & 0.7009 & 0.6975 \\
\end{array}
\]
\caption{True robust safety function $S^R(t,x)$.}
\end{table}

 The true robust safety function $S^R(t,x)$ is as follows:

\begin{table}[h!]
\centering
\label{tab:SR_values_symmetric}
\[
\begin{array}{c|cccc}
\text{time } t & S^R(t,1) & S^R(t,2) & S^R(t,3) & S^R(t,4) \\
\hline
0 & 0.8332 & 0.8332 & 0.8332 & 0.8332 \\
1 & 0.8332 & 0.8332 & 0.8332 & 0.8332 \\
2 & 0.8331 & 0.8331 & 0.8331 & 0.8331 \\
3 & 0.8327 & 0.8327 & 0.8327 & 0.8327 \\
4 & 0.8319 & 0.8319 & 0.8319 & 0.8319 \\
5 & 0.8299 & 0.8299 & 0.8299 & 0.8299 \\
6 & 0.8248 & 0.8248 & 0.8248 & 0.8248 \\
7 & 0.8120 & 0.8120 & 0.8120 & 0.8120 \\
8 & 0.7800 & 0.7800 & 0.7800 & 0.7800 \\
9 & 0.7000 & 0.7000 & 0.7000 & 0.7000 \\
\end{array}
\]
\caption{Robust safety function $S^R(t,x)$ with empirical average distribution.}
\end{table}

%So the robust safety function computed using the empirical distribution is sometimes lower than the true robust safety function. Thus, if we do not consider the ambiguity, quantified by $\rho$, introduced by finite-sample error, the true robust safety function can exceed the safety threshold $p$, while the robust safety function computed from the empirical distribution is lower than $p$. This, however, is not the case when we consider both $\delta$ and $\rho$ to compute the empirical robust safety function as shown in Table $1$.

Consequently, the robust safety function computed using only the empirical distribution can, in some cases, underestimate the true robust safety function. 
If the ambiguity arising from finite-sample uncertainty captured by $\rho$ is ignored, it is possible that the true robust safety function exceeds the safety threshold $p$, while the empirically computed robust safety function remains below $p$. This discrepancy does not occur when both sources of uncertainty, $\delta$ and $\rho$, are incorporated in the computation of the empirical robust safety function, as illustrated in Table~1. The empirical robust safety function always serves as an high confidence upper-bound for the true robust safety function.

\section{Conclusion and Future Work} \label{sec_conclusion}
We addressed the problem of distributionally robust safety verification for interval Markov Decision Processes (MDPs). Specifically, we studied the concept of robust probabilistic safety, termed robust $p$-safety, which generalizes probabilistic safety to MDPs with uncertain transition probabilities. We considered an ambiguity set that takes into account both model ambiguity and statistical (finite-sample) ambiguity. We provided a data-driven safety verification approach that estimates the probability of visiting unsafe states with an arbitrary high confidence. As future work, we will extend this framework to MDPs with larger state-action spaces and potentially continuous state spaces. 
%\cite{BUJORIANU2021665} in the framework of CPES.
\bibliography{IFAC_WC_ref} 
  % the appendices.
\end{document}